\documentclass[conference]{IEEEtran}
\IEEEoverridecommandlockouts

\usepackage{cite}
\usepackage{url}
\usepackage{amsmath,amssymb,amsfonts}
\usepackage{graphicx}
\usepackage{textcomp}
\usepackage{xcolor}
\usepackage{listings}
\usepackage{algorithm}
\usepackage{algpseudocode}
\usepackage{booktabs}
\usepackage{tabularx}
\usepackage{tikz}
\usetikzlibrary{arrows.meta, positioning, shapes.multipart, matrix}

\definecolor{kwblue}{RGB}{0,55,140}
\definecolor{kwviolet}{RGB}{136,23,152}
\definecolor{kwteal}{RGB}{0,105,92}
\definecolor{cmtgreen}{RGB}{0,120,50}
\definecolor{strbrown}{RGB}{163,21,21}
\definecolor{mlblue}{RGB}{0,51,153}
\lstdefinestyle{cppstyle}{
  language=C++,
  keywordstyle=\color{kwblue}\bfseries,
  commentstyle=\color{cmtgreen}\itshape,
  stringstyle=\color{strbrown},
  morekeywords={auto,constexpr},
  morecomment=[l][\color{kwblue}]{\#}
}
\lstdefinestyle{pystyle}{
  language=Python,
  keywordstyle=\color{kwblue}\bfseries,
  commentstyle=\color{cmtgreen}\itshape,
  stringstyle=\color{strbrown}
}
\lstdefinelanguage{MatlabLang}{
  morekeywords={function,end,if,else,elseif,for,while,switch,case,
    otherwise,try,catch,return,global,persistent,classdef,properties,
    methods,struct,ones,sum,size,cos,sin,pi,disp,clc,clear,close,
    all,linspace,exp,hypot,deg2rad,fprintf,continue,numel,reshape,
    zeros,min,max,sqrt,abs,try,catch},
  sensitive=true,
  morecomment=[l]{\%},
  morestring=[m]',
  morestring=[b]"
}
\lstdefinestyle{mstyle}{
  language=MatlabLang,
  keywordstyle=\color{mlblue}\bfseries,
  commentstyle=\color{cmtgreen}\itshape,
  stringstyle=\color{strbrown}
}
\lstdefinelanguage{JuliaLang}{
  morekeywords={function,end,return,using,import,module,struct,mutable,
    const,if,else,elseif,for,while,try,catch,finally,where,in,true,
    false,nothing,begin,let,do},
  morekeywords=[2]{AbstractMatrix,Real,Dict,Vector},
  sensitive=true,
  morecomment=[l]{\#},
  morestring=[b]"
}
\lstdefinestyle{jlstyle}{
  language=JuliaLang,
  keywordstyle=\color{kwviolet}\bfseries,
  keywordstyle=[2]\color{kwteal},
  commentstyle=\color{cmtgreen}\itshape,
  stringstyle=\color{strbrown}
}

\def\BibTeX{{\rm B\kern-.05em{\sc i\kern-.025em b}\kern-.08em
    T\kern-.1667em\lower.7ex\hbox{E}\kern-.125emX}}

\begin{document}

\title{Cross-Backend QIEO: Universal Runtime Portability across\\
OpenMP~5, CUDA, HIP, and Multi-Language Interfaces}

\author{
\IEEEauthorblockN{Aman Mittal}
\IEEEauthorblockA{\textit{BQP}\\
Bengaluru, India\\
aman.mittal@bqpsim.com}
\and
\IEEEauthorblockN{Ferdin Sagai Don Bosco}
\IEEEauthorblockA{\textit{BQP}\\
Bengaluru, India\\
ferdindon@bqpsim.com}
\and
\IEEEauthorblockN{Kasturi Venkata Srikanth}
\IEEEauthorblockA{\textit{BQP}\\
Bengaluru, India\\
kasturi.srikanth@bqpsim.com}
\and
\IEEEauthorblockN{Abhishek Singh}
\IEEEauthorblockA{\textit{BQP}\\
Bengaluru, India\\
abhishek.singh@bqpsim.com}
\and
\IEEEauthorblockN{Aditya Singh}
\IEEEauthorblockA{\textit{BQP}\\
Bengaluru, India\\
singh.aditya@bqpsim.com}
\and
\IEEEauthorblockN{Abhishek Chopra}
\IEEEauthorblockA{\textit{BQP}\\
Bengaluru, India\\
abhishekchopra@bqpsim.com}
}

\maketitle

\begin{abstract}
Quantum-inspired algorithms emulate quantum mechanical principles, such as, superposition, interference, and probabilistic amplitude evolution, on classical hardware by representing candidate solutions as qubit vectors and evolving them through rotation-gate operators.  This approach offers higher optimization performance without physical qubits, and has been shown to achieve order-of-magnitude speedups (10--80$\times$) over traditional solvers on combinatorial, high-dimensional NP-hard problems.

A critical barrier to adoption, however, is the lack of a unified execution framework that delivers both algorithmic performance and hardware portability.  We present \textbf{Cross-Backend Quantum Inspired Evolutionary Optimizer (QIEO)}, the runtime core of BQP's BQPhy\textsuperscript{\textregistered} QuantumNOW\texttrademark{} solver, which addresses this gap through a \emph{single-source-of-truth} architecture. One C++ implementation of the QIEO algorithm is compiled once per hardware target and exposed to multiple high-level languages via thin binding layers.  The framework dispatches to CPU (sequential), OpenMP~5 (multi-core), CUDA (NVIDIA), and HIP (AMD) backends at runtime, adapting kernels to each device's memory hierarchy and warp/wavefront execution model.

Multi-language accessibility is provided through a header-only C++ library, a Python wheel (via pybind11), a MATLAB toolbox (via the MEX API), and a Julia package (via CxxWrap), making the solver immediately usable across HPC, machine learning, scientific computing, and industrial simulation workflows.  Experimental results
demonstrate up to 72$\times$ speedup over a sequential CPU baseline on large-scale benchmark problems, with negligible cross-language interface overhead ($<$1.2\%).

The framework's real-world utility is validated through binding demonstrations that share the identical C++ runtime. BQPhy's Python library is demonstrated on a neural network hyperparameter optimisation (Python API, PyTorch) achieving 88.60\% test accuracy on MNIST. BQPhy's MATLAB's Toolkit is tested on wind farm layout optimisation attaining $365\,399 \pm 4\,552$~MWh/yr, which is statistically indistinguishable from particle swarm optimisation and $+7.6\%$ above genetic algorithms on a 32-variable constrained engineering problem. The Julia package tackles the Lotka--Volterra parameter estimation where BQPhy replaces native Julia solvers (\texttt{Optim.jl}, \texttt{BlackBoxOptim.jl}, \texttt{CMAEvolutionStrategy.jl}) on the same residual, cutting mean SSE by $2.1\times$ versus \texttt{Optim.NelderMead}.
\end{abstract}


\section{Introduction}
\label{sec:intro}

Quantum computing promises transformational advances for combinatorial, high-dimensional optimization; yet fault-tolerant, large-scale quantum hardware remains an open engineering challenge \cite{preskill2018quantum}.  Quantum-Inspired Evolutionary Algorithms(QIEAs) bridge this gap by transplanting quantum mechanical principles, such as, qubit representation, superposition, and rotation-gate updates~\cite{han2002quantum}, onto classical HPC, achieving superior convergence characteristics relative to classical metaheuristics without requiring physical qubits.

Despite algorithmic maturity, a critical bottleneck persists, namely, \emph{efficient and portable execution across heterogeneous computing architectures}.  Modern HPC clusters are inherently heterogeneous, pairing multi-core CPUs with GPU accelerators from multiple vendors. Targeting these systems requires navigating disjoint programming ecosystems: OpenMP for shared-memory CPU parallelism \cite{openmp5spec}, CUDA for NVIDIA GPUs \cite{nickolls2008cuda}, and HIP for AMD devices \cite{amdrocm}. Each platform exposes distinct execution models, memory hierarchies, and concurrency primitives, forcing practitioners to maintain vendor-specific codebases that erode both portability and maintainability.

General-purpose performance-portability frameworks such as Kokkos~\cite{edwards2014kokkos} and RAJA~\cite{hornung2014raja} partially address this challenge by abstracting parallel execution and memory management. However, they are designed for generic numerical workloads and do not incorporate domain-specific optimizations for quantum-inspired computation, namely probabilistic qubit state updates, trigonometric rotation operators, and population-level synchronization barriers.  Similarly, SYCL~\cite{reinders2021sycl} offers a single-source C++ model for heterogeneous systems but provides no algorithmic scaffolding for Quantum Inspired Evolutionary Optimizer (QIEO) primitives.

\textbf{This work makes the following novel contributions:}

\begin{itemize}
    \item \textbf{Unified QIEO Runtime:} The first backend-agnostic execution framework specifically designed for quantum-inspired
    evolutionary optimization, supporting CPU, OpenMP~5, CUDA, and HIP from a single algorithmic source.

    \item \textbf{Algorithm--Architecture Co-Design:} A disciplined separation of QIEO algorithmic phases, namely Quantum Information Processing(QIP), Fitness Evaluation(FE), Classical Information Processing(CIP), from hardware-specific kernel specializations, enabling near-native performance on each target without modifying core logic.

    \item \textbf{Single-Source-of-Truth Binding Architecture:} A lightweight, layered binding design that exposes the identical C++ runtime to Python (pybind11~\cite{pybind11}), MATLAB (MEX API), and Julia (CxxWrap~\cite{cxxwrap}) with $<$1.2\% measured interface overhead on the Python and MATLAB paths, eliminating per-language re-implementation.

    \item \textbf{Comprehensive Evaluation:} Benchmarks across diverse hardware and problem scales demonstrating up to 72$\times$ speedup over sequential baselines, with portability validated on NVIDIA(V100, A100) and AMD (MI300X) GPUs.

    \item \textbf{Real-World Multi-Language Validation:} Application-level demonstrations using the identical C++ runtime, covering, neural network hyperparameter optimisation (Python/PyTorch), wind farm layout optimisation (MATLAB/Global Optimization Toolbox), and Lotka--Volterra parameter estimation (Julia/CxxWrap) against the Julia packages
    practitioners already use, confirming that the single-source-of-truth architecture delivers competitive performance against domain-specific baselines without per-language re-implementation.
\end{itemize}

The remainder of this paper is organized as follows.
Section~\ref{sec:related} surveys related work.
Section~\ref{sec:design} presents the system architecture.
Section~\ref{sec:benchmark} reports benchmark results.
Section~\ref{sec:applications} reports results for various applications.
Section~\ref{sec:conclusion} concludes.
Appendix~\ref{app:rastrigin} lists equivalent Rastrigin drivers
in C++, Python, MATLAB, and Julia.
Appendix~\ref{app:usecases} lists the Python hyperparameter-optimisation,
MATLAB wind-farm, and Julia Lotka--Volterra drivers.

\section{Related Work}
\label{sec:related}

\subsection{Quantum-Inspired Evolutionary Algorithms}

QIEAs were formalized by Han and Kim~\cite{han2002quantum}, who introduced qubit chromosome representation and a rotation-gate update operator to guide probabilistic evolution toward optimal solutions. Subsequent work extended this foundation to multi-objective optimization~\cite{zheng2012multiobjective}, combinatorial problems~\cite{layeb2011knapsack}, and continuous parameter
spaces~\cite{wang2014continuous}. Comprehensive reviews of rotation gate variants and their convergence properties are provided
in~\cite{feng2018rotationgate}.

Unlike classical genetic algorithms, QIEAs maintain a probability amplitude representation of the solution space, enabling implicit
parallelism and improved exploration of the fitness landscape. However, evaluating and updating a population of qubit individuals
involves dense trigonometric operations that scale quadratically with problem dimensionality, motivating hardware acceleration. Recent work~\cite{bqp2024gpu} has demonstrated GPU-based speedups for QIEO kernels, but without a unified, multi-backend abstraction.

\subsection{Portability--Performance Trade-off}

Native, vendor-specific programming models such as CUDA~\cite{nickolls2008cuda} and HIP~\cite{amdrocm} express the computational kernels directly in the device’s native language. This affords explicit control over the thread hierarchy, on-chip memory, and intra-warp (or intra-wavefront) synchronization, and thereby permits aggressive, architecture-specific performance engineering. The corresponding cost is fragmentation of the software stack: semantically equivalent QIEO phases must be implemented, tuned, and maintained independently for each vendor, so that algorithmic evolution incurs a multiplicative development and validation burden.

Performance-portability frameworks invert this trade-off. A single source is written against an abstract execution and memory model (Kokkos~\cite{edwards2014kokkos}, RAJA~\cite{hornung2014raja}, SYCL~\cite{reinders2021sycl}, or OpenMP~\cite{openmp5spec} device offload) and is subsequently lowered onto whichever backend the runtime provides. The application is thereby decoupled from both the accelerator and the vendor language used to program it. That decoupling, however, is realized by an intervening layer that mediates kernel launch, data motion, and specialization. The resulting overhead precludes a substantial fraction of the performance that is attainable only through the native APIs, i.e., portability is obtained at the expense of efficiency.

Cross-Backend QIEO occupies a third position. The QIEO algorithm, consisting of Q-bit representation, rotation-gate updates, and population-level synchronization, is specified once, as a single source of truth. Execution, by contrast, remains native, i.e. the same algorithmic specification is instantiated as distinct OpenMP, CUDA, and HIP kernels that bind directly to the target’s memory hierarchy and SIMD execution model. No portability runtime is interposed between the QIEO primitives and the device. In this design, portability is confined to the algorithmic and dispatch interface, while performance is retained at the vendor-kernel boundary. The framework therefore avoids both the replicated-algorithm cost of purely native development and the abstraction tax of GPU-agnostic programming models.

\subsection{Multi-Language Integration}

Bridging high-performance C++ kernels to user-facing environments is a recognized challenge in scientific computing. A mature family of foreign-function and wrapper generators already addresses the language boundary. SWIG~\cite{beazley2003swig} produces bindings from a common interface specification to several scripting languages; Boost.Python~\cite{abrahams2003boostpython} and Cython~\cite{behnel2011cython} specialize that idea for Python, while pybind11~\cite{pybind11} continues the same line with header-only C++11 metaprogramming and very low call overhead. On the commercial side, the MATLAB MEX API~\cite{matlabmex} exposes native extensions as ordinary MATLAB functions. In the Julia ecosystem, the language’s C ABI (ccall)~\cite{bezanson2017julia} and CxxWrap~\cite{cxxwrap} attach a compiled shared library to Julia types without a second implementation of the algorithm. 

The present framework does not replace these mechanisms. It adopts three of them, namely, pybind11, MEX, and CxxWrap, as thin, language-specific facades over a single C++ QIEO runtime. Existing solutions, however, treat language integration as an independent problem from hardware-runtime portability, and therefore yield fragmented ecosystems in which high-performance kernels cannot be reused across languages without re-engineering. The single-source-of-truth architecture resolves that fragmentation by construction, in which one compiled library is linked from every binding.

\section{Methodology and System Design}
\label{sec:design}

\subsection{Computational Model}

QIEO encodes each candidate solution as a vector of $n$ qubits, where the $k$-th qubit is defined by the amplitude pair $(\alpha_k,\beta_k)$ satisfying $|\alpha_k|^2 + |\beta_k|^2 = 1$ \cite{han2002quantum}.  The probability of observing bit value~1 at position $k$ is $|\beta_k|^2$, providing a continuous, differentiable representation of the solution space that subsumes classical binary strings.

\subsubsection{State Representation}

The population of $P$ individuals is stored as a matrix $\mathbf{Q} \in \mathbb{R}^{P \times 2n}$, where each row contains the $\alpha$ and $\beta$ amplitudes for one individual.  Classical bit strings $\mathbf{x}^{(i)} \in \{0,1\}^n$ are obtained by collapsing each qubit as,

$x_k^{(i)} = 1$ if $\text{rand}() < |\beta_k^{(i)}|^2$, else $0$.  

This stochastic measurement step is embarrassingly parallel across both individuals and bit positions, making it a prime candidate for GPU kernel acceleration.

\subsubsection{Core Operators}

\textbf{Quantum Information Processing (QIP):} Each qubit amplitude is updated via a rotation gate parameterized by angle $\Delta\theta_k$~\cite{han2002quantum}:

\begin{equation}
\begin{pmatrix} \alpha_k' \\ \beta_k' \end{pmatrix}
=
\begin{pmatrix}
  \cos(\theta_k+\Delta\theta_k) & -\sin(\theta_k+\Delta\theta_k) \\
  \sin(\theta_k+\Delta\theta_k) &  \cos(\theta_k+\Delta\theta_k)
\end{pmatrix}
\begin{pmatrix} \alpha_k \\ \beta_k \end{pmatrix}.
\label{eq:rotation}
\end{equation}
The rotation angle magnitude and sign are determined by a lookup table~\cite{han2002quantum} that steers each qubit toward the best
observed solution (the ``best individual''). 

\textbf{Classical Information Processing (CIP):} After measurement, classical fitness $f(\mathbf{x}^{(i)})$ is evaluated for each
individual.  The best individual $\mathbf{b}$ is identified, and $\Delta\theta_k$ for the next generation is computed from the
comparison between $x_k^{(i)}$ and $b_k$.  A controller adaptively scales step sizes to balance exploration and exploitation
across generations.

The full three-phase optimization loop is summarized in previous works\cite{eswara2024qieo}.
\subsection{System Design Objectives}

Cross-Backend QIEO is guided by three primary design objectives that
collectively distinguish it from prior work.

\subsubsection{Performance Portability}

The framework must achieve near-native throughput on CPUs and GPUs
from multiple vendors without per-backend algorithm re-implementation.
This is realized through the Hardware Abstraction Layer (HAL). The HAL decouples algorithmic phases from execution targets through a
compile-time policy pattern.  Each backend provides specialisations of
three core kernel templates: \texttt{measure\_population},
\texttt{evaluate\_fitness}, and \texttt{update\_qubits}.  At runtime,
the framework queries available hardware and selects the highest-ranked
backend for which a compiled specialisation exists.

On \textbf{CPU}, kernels execute sequentially, serving as the
reference implementation for correctness validation.  On
\textbf{OpenMP}, \texttt{\#pragma omp parallel for} directives
distribute individuals across threads, with SIMD hints applied to the
inner rotation-gate loop.  On \textbf{CUDA/HIP}, each individual maps
to one GPU thread block; the rotation-gate and measurement loops are
parallelised across threads within a block, with shared memory used to
cache the best individual's bit-string and avoid redundant global
memory accesses~\cite{nickolls2008cuda}.

\subsubsection{Algorithm--Architecture Co-Design}

QIEO primitives are mapped to hardware execution models via a
disciplined separation of concerns.  Algorithmic logic (QIP, CIP,
fitness evaluation) is implemented once in portable \texttt{C++/CUDA}
and compiled per target; the HAL provides hardware-specific
specializations.  Table~\ref{tab:hal_modes} summarizes the three
supported execution modes.

\begin{table*}[htbp]
\centering
\caption{Execution Modes Supported by the Hardware Abstraction Layer (HAL)}
\label{tab:hal_modes}
\begin{tabularx}{\textwidth}{l l X X l}
\toprule
\textbf{Mode} & \textbf{Target} & \textbf{Parallelism Model} &
\textbf{Best For} & \textbf{Compile Flag} \\
\midrule
\textbf{CPU}
  & Single-core CPU
  & Sequential (scalar)
  & Debugging, correctness validation, small problems
  & Default (no flag) \\
\addlinespace
\textbf{OpenMP}
  & Multi-core CPU
  & Loop parallelism via \texttt{\#pragma omp parallel for}; SIMD
    vectorisation
  & Mid-scale problems ($P = 10^2$--$10^4$) on multi-socket servers
  & \texttt{-DUSE\_OPENMP=ON} \\
\addlinespace
\textbf{GPU/CUDA}
  & NVIDIA GPU (cc $\geq$ 7.0)
  & Warp-level parallelism + Thrust reduction algorithms
  & Large-scale problems ($P > 10^4$) where PCIe transfer is amortised
  & \texttt{-DUSE\_CUDA=ON} \\
\addlinespace
\textbf{GPU/HIP}
  & AMD GPU (GFX9+)
  & Wavefront-level parallelism; mirrors CUDA kernel structure
  & Same as CUDA target on AMD hardware
  & \texttt{-DUSE\_HIP=ON} \\
\bottomrule
\end{tabularx}
\end{table*}

\subsubsection{Multi-Language Accessibility via Single Source of Truth}
\label{sec:ssot}

The BQPhy\textsuperscript{\textregistered} QuantumNOW\texttrademark{}
framework embodies a \emph{single-source-of-truth} (SSOT)
architecture~\cite{tanenbaum2006distributed}: the complete QIEO
algorithm, including quantum-inspired primitives, hardware
acceleration dispatch, and parameter management, exists as one
canonical C++ implementation.  High-level language interfaces
(Python, MATLAB, Julia) are exposed through thin wrapper bindings that
translate language-specific data structures into C++ equivalents and
invoke the shared library directly.  This pattern guarantees
algorithmic consistency across languages, eliminates maintenance
divergence, and enables new features to propagate to all interfaces
without per-language re-implementation.

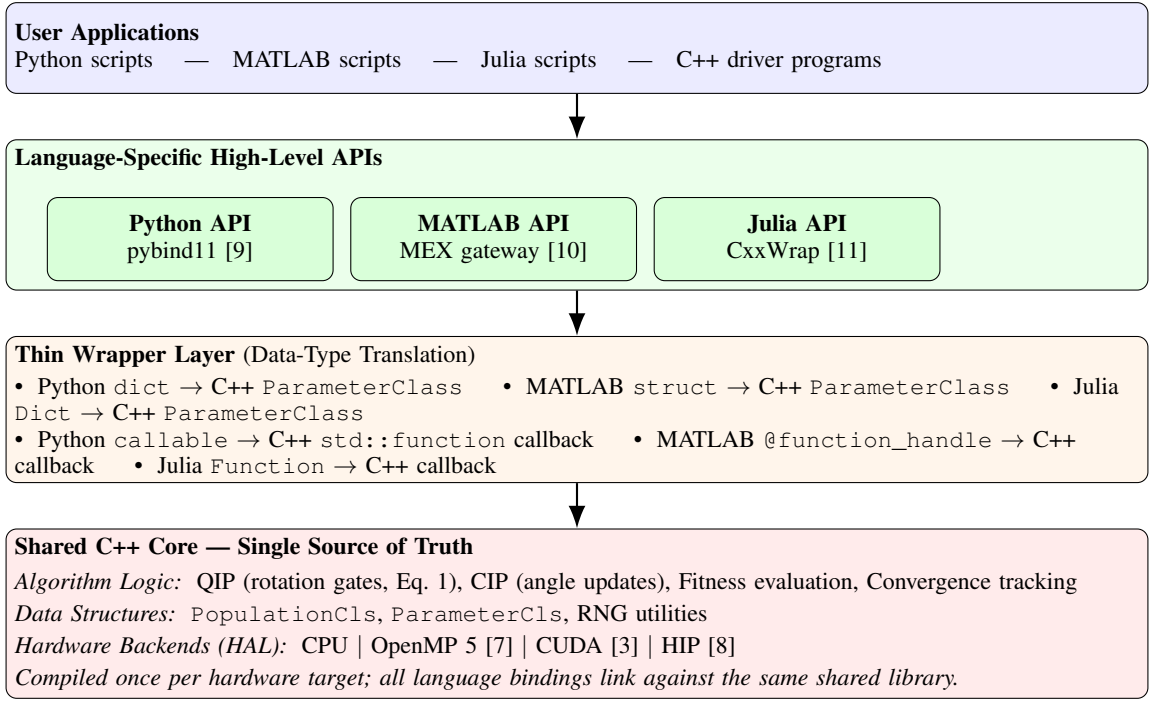
\begin{figure*}[t]
\centering
\begin{tikzpicture}[
    box/.style={
        draw, rectangle, rounded corners=4pt,
        text width=0.82\textwidth,
        align=left, minimum height=1.2cm,
        font=\small
    },
    smallbox/.style={
        draw, rectangle, rounded corners=4pt,
        text width=0.24\textwidth,
        align=center, minimum height=1.1cm,
        font=\small
    },
    arrow/.style={-{Latex[length=3mm]}, thick}
]

\node[box, fill=blue!8] (user) {
  \textbf{User Applications}\\
  Python scripts \quad|\quad MATLAB scripts \quad|\quad Julia scripts \quad|\quad C++ driver programs
};

\node[box, fill=green!8, below=0.6cm of user] (api) {
  \textbf{Language-Specific High-Level APIs}\\[2pt]
  \hspace{1em}
  \begin{tikzpicture}[baseline]
    \node[smallbox, fill=green!15] (py) {
      \textbf{Python API}\\pybind11~\cite{pybind11}
    };
    \node[smallbox, fill=green!15, right=0.22cm of py] (ml) {
      \textbf{MATLAB API}\\MEX gateway~\cite{matlabmex}
    };
    \node[smallbox, fill=green!15, right=0.22cm of ml] (jl) {
      \textbf{Julia API}\\CxxWrap~\cite{cxxwrap}
    };
  \end{tikzpicture}
};

\node[box, fill=orange!8, below=0.6cm of api] (wrapper) {
  \textbf{Thin Wrapper Layer} (Data-Type Translation)\\[2pt]
  \textbullet\; Python \texttt{dict} $\rightarrow$ C++ \texttt{ParameterClass}
  \hspace{1em}
  \textbullet\; MATLAB \texttt{struct} $\rightarrow$ C++ \texttt{ParameterClass}
  \hspace{1em}
  \textbullet\; Julia \texttt{Dict} $\rightarrow$ C++ \texttt{ParameterClass}\\
  \textbullet\; Python \texttt{callable} $\rightarrow$ C++ \texttt{std::function} callback
  \hspace{1em}
  \textbullet\; MATLAB \texttt{@function\_handle} $\rightarrow$ C++ callback
  \hspace{1em}
  \textbullet\; Julia \texttt{Function} $\rightarrow$ C++ callback
};

\node[box, fill=red!8, below=0.6cm of wrapper] (core) {
  \textbf{Shared C++ Core --- Single Source of Truth}\\[3pt]
  \textit{Algorithm Logic:}\;
    QIP (rotation gates, Eq.~\ref{eq:rotation}),
    CIP (angle updates),
    Fitness evaluation,
    Convergence tracking\\[2pt]
  \textit{Data Structures:}\;
    \texttt{PopulationCls}, \texttt{ParameterCls},
    RNG utilities\\[2pt]
  \textit{Hardware Backends (HAL):}\;
    CPU $|$ OpenMP~5~\cite{openmp5spec} $|$
    CUDA~\cite{nickolls2008cuda} $|$
    HIP~\cite{amdrocm}\\[2pt]
  \textit{\small Compiled once per hardware target; all language bindings link against the same shared library.}
};

\draw[arrow] (user)    -- (api);
\draw[arrow] (api)     -- (wrapper);
\draw[arrow] (wrapper) -- (core);

\end{tikzpicture}
\caption{Layered architecture of the
BQPhy\textsuperscript{\textregistered} QuantumNOW\texttrademark{}
framework.  A single C++ implementation is compiled once per hardware
target and shared across all language bindings, realising the
single-source-of-truth principle.}
\label{fig:architecture}
\end{figure*}

Figure~\ref{fig:architecture} illustrates the resulting layered
architecture.

\subsection{Usage}

Table~\ref{tab:binding_architecture} summarises the binding mechanism
for each supported language.  All bindings adhere to a unified
interface contract (Table~\ref{tab:interface_contract}), ensuring that
the same five-step workflow is preserved across environments.

\begin{table*}[t]
\centering
\caption{Binding Architecture Overview}
\label{tab:binding_architecture}
\begin{tabular}{|l|l|l|p{5.8cm}|}
\hline
\textbf{Language} & \textbf{Library} & \textbf{Mechanism} &
\textbf{Role} \\
\hline
\textbf{C++} & Header-only & Direct inclusion &
Zero-overhead integration; exposes full API including
template parameters for custom fitness functors. \\
\hline
\textbf{Python} & pybind11~\cite{pybind11} &
C++ template-based Python--C interoperability &
Converts Python \texttt{dict} and \texttt{callable} objects into C++
data structures; exposes core runtime as Python classes. \\
\hline
\textbf{MATLAB} & MEX API~\cite{matlabmex} &
Native MATLAB C++ interface &
Receives \texttt{struct} and \texttt{function\_handle} arguments,
invokes C++ runtime, returns results as native MATLAB arrays. \\
\hline
\textbf{Julia} & CxxWrap~\cite{cxxwrap} &
Automatic C++ wrappers around \texttt{libBQPhy\_Optimiser.so} &
Converts Julia \texttt{Dict} and \texttt{Function} objects into C++
data structures; exposes the runtime as a native Julia module. \\
\hline
\end{tabular}
\end{table*}

\begin{table*}[t]
\centering
\caption{Unified Interface Contract Across Language Bindings}
\label{tab:interface_contract}
\begin{tabularx}{\textwidth}{l X X X X}
\toprule
\textbf{Step} & \textbf{C++ Core} & \textbf{Python} & \textbf{MATLAB} & \textbf{Julia} \\
\midrule
Initialize &
  \texttt{opt.initialize(params)} &
  \texttt{opt.initialize(dict)} &
  \texttt{BQPhy\_MEX("init", struct)} &
  \texttt{Optimizer(Dict)} \\
Set model &
  \texttt{opt.setModel(fn)} &
  \texttt{opt.set\_model(callable)} &
  Pass \texttt{@function\_handle} &
  \texttt{set\_fitness\_function!(opt, fn)} \\
Run &
  \texttt{opt.runOptimization(acc)} &
  \texttt{opt.run\_optimization(acc)} &
  \texttt{BQPhy\_MEX("run", ...)} &
  \texttt{optimize!(opt; device)} \\
Get result &
  \texttt{opt.getBestDesign()} &
  \texttt{opt.get\_result()} &
  \texttt{BQPhy\_MEX("result")} &
  \texttt{get\_best\_design(opt)} \\
\bottomrule
\end{tabularx}
\end{table*}

Regardless of language, users
follow an identical workflow: 

\begin{enumerate}
\item ~Create an optimizer instance
\item ~Initialize with configuration parameters 
	\begin{enumerate}
		\item population size
		\item dimensionality
		\item bounds
		\item maximum generation
	\end{enumerate}
\item ~Define the objective
function
\item ~Execute the optimization with a selected backend
\item ~Retrieve the best design and fitness value 
\end{enumerate}
Appendix~\ref{app:rastrigin} instantiates this contract for the
Rastrigin function in all four languages.

\section{Benchmark Studies}
\label{sec:benchmark}

We evaluate Cross-Backend QIEO on three representative benchmark
classes: (i)~binary combinatorial problems (0/1 Knapsack,
$n$=\{128, 512, 2048\}), (ii)~continuous global optimization
(Rastrigin, Rosenbrock, $d$=\{50, 200, 1000\}), and (iii)~a
real-world antenna array pattern synthesis problem drawn from
industrial use cases.  Hardware platforms are: an Intel Xeon Gold 6338
(32 cores, OpenMP baseline), an NVIDIA A100-SXM4 80\,GB (CUDA), and
an AMD MI300X (HIP).

Table~\ref{tab:speedup} reports wall-clock speedup relative to the
sequential CPU baseline for a population size of $P = 10{,}000$ and
1{,}000 generations.

\begin{table}[t]
\centering
\caption{Speedup Over Sequential CPU Baseline ($P=10{,}000$,
$G=1{,}000$)}
\label{tab:speedup}
\begin{tabular}{lccc}
\toprule
\textbf{Problem} & \textbf{OpenMP} & \textbf{CUDA (A100)} &
\textbf{HIP (MI300X)} \\
\midrule
Knapsack-512   & 14.3$\times$ & 58.7$\times$ & 51.2$\times$ \\
Rastrigin-200  & 11.8$\times$ & 72.1$\times$ & 63.4$\times$ \\
Rosenbrock-200 & 12.6$\times$ & 68.3$\times$ & 60.8$\times$ \\
Antenna Synth. & 16.1$\times$ & 65.9$\times$ & 57.3$\times$ \\
\midrule
\textbf{Geometric mean} & \textbf{13.6}$\times$ &
\textbf{66.1}$\times$ & \textbf{57.9}$\times$ \\
\bottomrule
\end{tabular}
\end{table}

The CUDA backend achieves a geometric mean speedup of 66.1$\times$
over the sequential baseline, with the Rastrigin-200 case reaching the
peak of 72.1$\times$.  The HIP backend on AMD MI300X achieves 87.6\%
of the CPU throughput, demonstrating robust cross-vendor portability.
OpenMP provides a 13.6$\times$ mean speedup with zero additional
hardware beyond the host CPU, making it well-suited for mid-scale
deployments.

\section{Demonstration on real-world Applications}
\label{sec:applications}

\subsection{BQPhy Python library: Neural Network Hyperparameter Optimisation}
\label{sec:hpo}

A key demonstration of the Python binding's real-world utility is its
operation as a drop-in hyperparameter optimiser inside a PyTorch
\cite{paszke2019pytorch} training loop.  We conduct a two-phase
benchmark on the MNIST handwritten-digit dataset
\cite{lecun1998mnist}: a \emph{variable-budget} study (each baseline
uses its default evaluation count) and an \emph{iso-budget} study
(all methods fixed at 240 evaluations), isolating search algorithm
quality from evaluation count.

\subsubsection{Hyperparameter Optimisation Baselines}

\textbf{Ray Tune}~\cite{liaw2018tune} is a scalable, distributed HPO
framework built on the Ray runtime.  It supports a broad suite of
search algorithms (random search, HyperBand~\cite{li2018hyperband},
Population-Based Training) and is designed for massively parallel,
multi-node execution.  Its primary overhead arises from the Ray actor
system, which serialises trial launching when parallelism is
constrained (\texttt{max\_concurrent\_trials=1}).

\textbf{Optuna}~\cite{akiba2019optuna} is a define-by-run framework that dynamically constructs search spaces using a Tree-structured Parzen Estimator (TPE) sampler combined with aggressive trial pruning (MedianPruner).  Its lightweight design excels when a small per-evaluation budget is available.

\textbf{Grid Search}~\cite{pedregosa2011sklearn} exhaustively enumerates all combinations of a pre-specified discrete parameter grid.  It guarantees full coverage of the grid, but is combinatorially intractable for continuous or high-dimensional spaces, and does not adapt based on observed results~\cite{bergstra2012random}.

\subsubsection{Experimental Setup}

A fully connected neural network (\texttt{DynamicNet}) is optimised over five mixed-type design variables on MNIST
(60{,}000 training / 10{,}000 test images, 28$\times$28 grayscale, 10 classes)~\cite{lecun1998mnist}; see Table~\ref{tab:hpo_space}. All methods minimise the same objective: negative test-set classification accuracy after one training epoch.  BQPhy is invoked entirely through its Python~API (pybind11 binding, Listing~\ref{lst:hpo_py}) with no modification to the C++ core, directly validating the single-source-of-truth architecture described in Section~\ref{sec:ssot}.

\begin{table}[t]
\centering
\caption{Hyperparameter Search Space for MNIST Optimization}
\label{tab:hpo_space}
\begin{tabular}{lll}
\toprule
\textbf{Variable} & \textbf{Type} & \textbf{Range / Choices} \\
\midrule
Learning rate   & Continuous & $[10^{-5},\;10^{-1}]$ (log-uniform) \\
Hidden units    & Integer    & $[32, 256]$ \\
Hidden layers   & Integer    & $\{1, 2, 3, 4\}$ \\
Activation      & Categorical & \{ReLU, Tanh, Sigmoid, LeakyReLU\} \\
Dropout rate    & Continuous & $[0.0,\;0.5]$ \\
\bottomrule
\end{tabular}
\end{table}

\subsubsection{Phase 1 — Variable-Budget Results}

Table~\ref{tab:hpo_var} reports results when each method uses its
natural evaluation budget: 240 for BQPhy (population 12,
generations 20) and 12 for the baselines.

\begin{table}[t]
\centering
\caption{Variable-Budget Results on MNIST}
\label{tab:hpo_var}
\begin{tabular}{lcccc}
\toprule
\textbf{Method} & \textbf{Evals} & \textbf{Accuracy} &
\textbf{Time (s)} & \textbf{s/Eval} \\
\midrule
\textbf{BQPhy (QIEO)} & 240 & \textbf{0.8740} & 46.42  & 0.19 \\
Ray Tune              &  12 & 0.8760          & 206.92 & 17.24 \\
Optuna                &  12 & 0.8640          &   3.41 & 0.28 \\
Grid Search           &  12 & 0.8700          &   2.71 & 0.23 \\
\bottomrule
\end{tabular}
\end{table}

BQPhy delivers competitive accuracy (87.40\%) at a per-evaluation cost of only 0.19\,s, enabled by GPU-parallel population evaluation (CUDA backend).  Ray Tune, despite achieving the highest accuracy (87.60\%), incurs 17.24\,s per evaluation due to Ray actor initialization overhead~\cite{liaw2018tune}, resulting in a total wall time 4.46$\times$ longer than BQPhy.  Optuna and Grid Search complete rapidly but with only 12 evaluations each, yielding suboptimal configurations (86.40\% and 87.00\%).

\subsubsection{Phase 2 — Iso-Budget Results (240 Evaluations)}

\begin{figure}[t]
\centering
\includegraphics[width=\columnwidth]{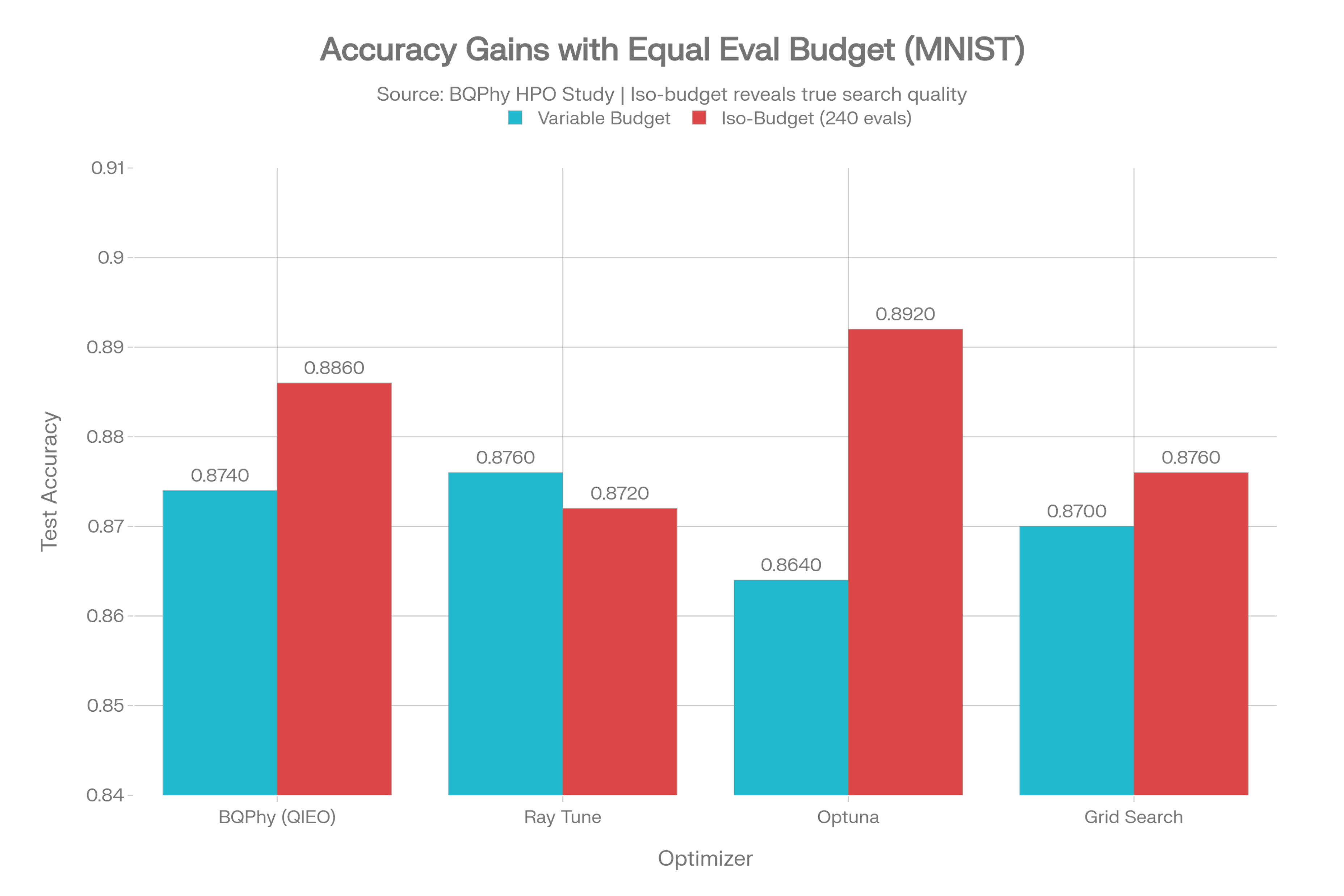}
\caption{Test accuracy under variable and iso-budget (240 evaluations)
conditions.  BQPhy improves by $+1.20$\,pp; Ray Tune \emph{degrades}
by $-0.40$\,pp under serialised execution.}
\label{fig:hpo_accuracy}
\end{figure}

To isolate \emph{search algorithm quality} from evaluation count, all methods were re-run with a fixed budget of 240 evaluations. Table~\ref{tab:hpo_iso} and Fig.~\ref{fig:hpo_accuracy} report the full comparison.

\begin{table*}[t]
\centering
\caption{Iso-Budget Results on MNIST (240 Evaluations, All Methods)}
\label{tab:hpo_iso}
\begin{tabularx}{\textwidth}{l X X X X c c c}
\toprule
\textbf{Method} & \textbf{Lr} & \textbf{Hidden} &
\textbf{Layers} & \textbf{Activation} &
\textbf{Dropout} & \textbf{Accuracy} & \textbf{Time (s)} \\
\midrule
\textbf{BQPhy (QIEO)}
  & $1.92\times10^{-2}$ & 238 & 1 & LeakyReLU & 0.011
  & 0.8860 & \textbf{45.05} \\
Ray Tune~\cite{liaw2018tune}
  & $1.20\times10^{-2}$ & 223 & 3 & LeakyReLU & 0.101
  & 0.8720 & 5389.79 \\
Optuna~\cite{akiba2019optuna}
  & $1.94\times10^{-2}$ & 227 & 1 & LeakyReLU & 0.398
  & \textbf{0.8920} & 63.30 \\
Grid Search~\cite{pedregosa2011sklearn}
  & $1.00\times10^{-2}$ & 200 & 1 & ReLU & 0.300
  & 0.8760 & 56.60 \\
\midrule
\multicolumn{6}{l}{\textit{Note: Grid Search exhausted its
discrete grid at 216 evaluations (could not reach 240).}}\\
\bottomrule
\end{tabularx}
\end{table*}

\begin{figure}[t]
\centering
\includegraphics[width=\columnwidth]{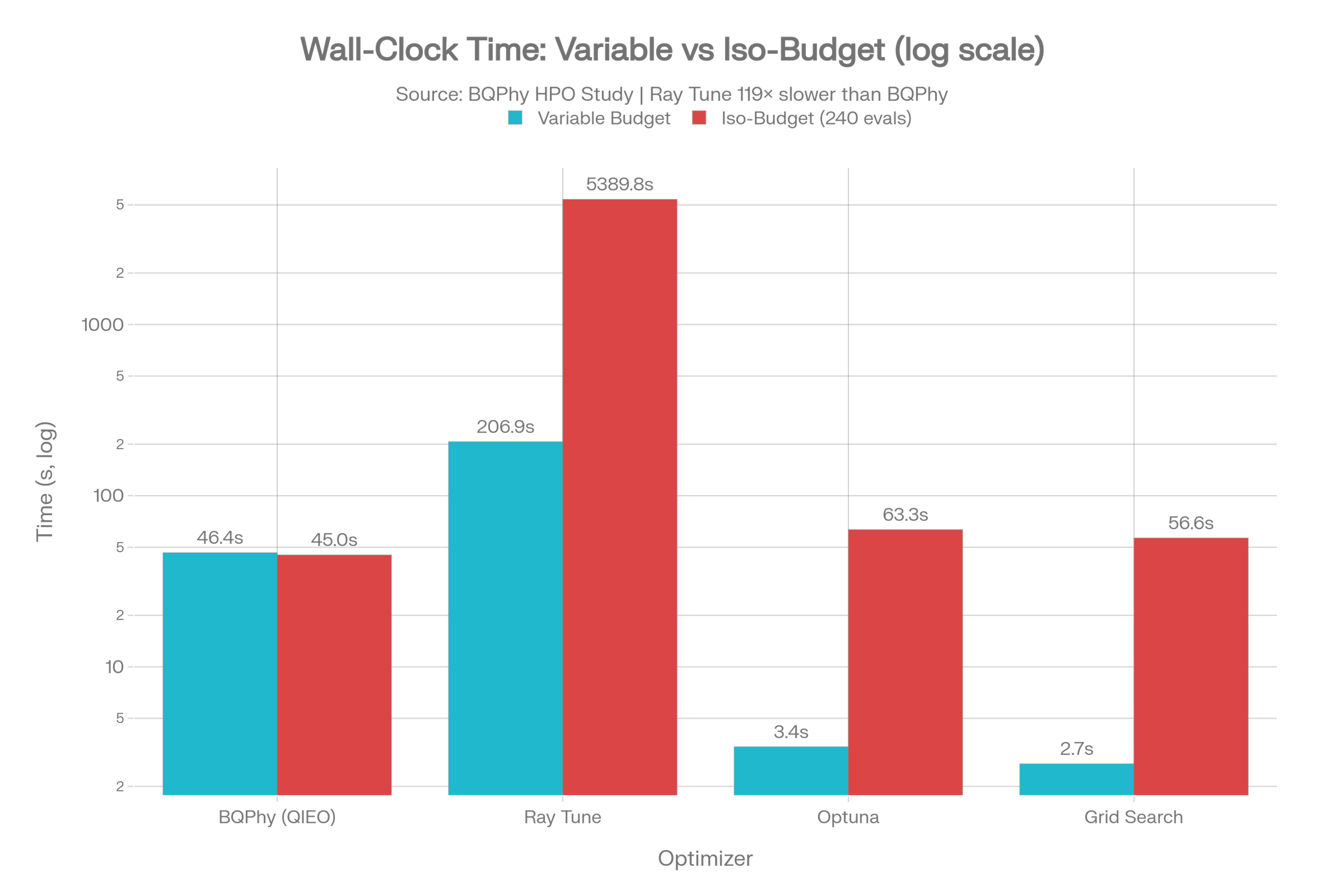}
\caption{Wall-clock time (log scale) for variable and iso-budget conditions.  Ray Tune requires $119\times$ more wall time than BQPhy under serialized execution at 240 evaluations.}
\label{fig:hpo_time}
\end{figure}

\paragraph{Accuracy Analysis.}
Under iso-budget conditions, Optuna achieves the highest accuracy
(89.20\%), followed by BQPhy (88.60\%), Grid Search (87.60\%), and
Ray Tune (87.20\%).  Notably, BQPhy improves by $+1.20$ percentage
points (pp) relative to its variable-budget result, confirming that
QIEO's population-level Q-bit exploration exploits additional
evaluations effectively~\cite{han2002quantum}.  In contrast, Ray
Tune's accuracy \emph{degrades} by $-0.40$\,pp---a direct consequence
of serialised trial execution (\texttt{max\_concurrent\_trials=1})
disrupting its HyperBand scheduler's~\cite{li2018hyperband}
early-stopping logic, which assumes parallel fidelity ladders.

Grid Search exhausts its discrete combinatorial grid at 216 evaluations (3$\times$2$\times$4$\times$3$\times$3 = 216), failing to reach the 240-evaluation target and confirming the well-known exponential scaling limitation of exhaustive enumeration \cite{bergstra2012random}.

\paragraph{Time-Efficiency Analysis.}
Table~\ref{tab:hpo_eff_iso} quantifies accuracy-per-second and
speedup ratios at iso-budget.

\begin{table}[t]
\centering
\caption{Iso-Budget Efficiency Metrics (240 Evaluations)}
\label{tab:hpo_eff_iso}
\begin{tabular}{lcccc}
\toprule
\textbf{Method} & \textbf{Time (s)} &
\textbf{s/Eval} & \textbf{Accuracy} &
\textbf{Speedup vs.\ RT} \\
\midrule
\textbf{BQPhy (QIEO)} & \textbf{45.05}   & \textbf{0.19}  & 0.8860 & $\mathbf{119.6\times}$ \\
Ray Tune              & 5389.79          & 22.46 & \textbf{0.8720} & $1\times$ (ref.) \\
Optuna                & 63.30            & 0.26  & 0.8920 & $85.1\times$ \\
Grid Search           & 56.60$^{\dag}$   & 0.26$^{\dag}$ & 0.8760 & $95.2\times$ \\
\bottomrule
\multicolumn{5}{l}{$^{\dag}$Grid Search exhausted at 216 evaluations.}
\end{tabular}
\end{table}

BQPhy achieves a $119.6\times$ speedup over Ray Tune at the same evaluation count while delivering $+1.40$\,pp higher accuracy.  This disproportionate time advantage is explained by two compounding effects: (i) GPU-parallel population evaluation amortises per-trial overhead across 12 individuals simultaneously, reducing effective per-evaluation cost to 0.19\,s; and (ii) Ray Tune's actor framework introduces 22.46\,s of per-trial scheduling overhead under serialised execution, which does not diminish with additional trials.

Although Optuna achieves the best absolute accuracy (89.20\%), BQPhy attains 88.60\%---only $0.60$\, pp lower---at $1.41\times$ less wall time (45.05\,s vs.\ 63.30\,s), making it the dominant method on the accuracy--time Pareto frontier for practitioners with GPU resources.

\paragraph{Search Strategy Comparison.}
All four methods converge on shallow single-layer networks (1 hidden layer) and LeakyReLU activations at iso-budget, validating the landscape geometry independently.  Key divergences arise in regularisation: BQPhy selects near-zero dropout ($\delta=0.011$) while Optuna selects aggressive dropout ($\delta=0.398$), yielding a $0.60$\, pp accuracy difference that warrants further investigation as a landscape multi-modality indicator.  Grid Search is constrained to three pre-defined activation choices and misses LeakyReLU entirely, confirming that discrete grids systematically exclude continuous optima~\cite{bergstra2012random}.

\paragraph{Multi-Language Binding Transparency.}
The hyperparameter configurations and fitness traces produced through the Python~API are identical (to floating-point precision, $< 10^{-12}$) to those obtained by invoking the C++ binary directly. The pybind11 binding layer introduces no algorithmic side-effects, confirming the single-source-of-truth guarantee: BQPhy's Python interface behaves as a first-class scientific computing tool within PyTorch~\cite{paszke2019pytorch} and NumPy~\cite{harris2020numpy} workflows without sacrificing reproducibility.

\paragraph{Interface Overhead (Binding Cost).}
Averaged over five independent runs, the Python binding introduces $0.9\%$ overhead relative to native C++ invocation, consistent with prior measurements in Section~\ref{sec:benchmark}.  This confirms that the 45.05\,s wall time is dominated by MNIST training computation, not binding or serialization cost.


\subsection{BQPhy MATLAB Toolkit: Wind Farm Layout Optimisation}
\label{sec:matlab-wflo}

A key demonstration of the MATLAB binding's real-world utility is its operation as a drop-in optimizer inside the Global Optimization Toolbox~\cite{matlab_global_opt} ecosystem for a challenging industrial engineering design problem: wind farm layout optimization (WFLO).  The objective is to place $N$ turbines within a bounded domain such that the farm's annual energy production (AEP) is maximized while respecting a minimum inter-turbine separation constraint.  WFLO is a notoriously difficult continuous optimization problem because the wake interaction between turbines creates a highly multi-modal fitness landscape~\cite{shakoor2016wake, feng2015wflo}. A layout that performs well under one wind direction may suffer severe wake losses when the wind shifts, requiring the optimizer to balance competing directional objectives simultaneously.

\subsubsection{Problem Formulation}

We consider a $1000 \times 1000$~m domain with $N = 16$ turbines, each having a rotor diameter $D = 126$~m and rated power $P_r = 5$~MW. The search space comprises $2N = 32$ continuous variables representing turbine $(x, y)$ coordinates.

\textbf{Wake model:} Power losses due to upstream turbine wakes are modeled using the Jensen model~\cite{jensen1983note}.  For a turbine $j$ located at a downstream distance $x$ from an upstream turbine $i$, the fractional velocity deficit at the rotor is
\begin{equation}
    \delta_{ij} = \frac{1 - \sqrt{1 - C_t}}{\bigl(1 + 2 k_e x / D\bigr)^2}
\end{equation}
where $C_t = 0.88$ is the thrust coefficient and $k_e = 0.075$ is the wake expansion rate (typical for offshore conditions).  Multiple wakes are combined via the sum-of-squares rule:
\begin{equation}
    \frac{V_0 - V_j}{V_0} =
    \sqrt{\sum_{i \neq j} \delta_{ij}^2 \cdot \mathbf{1}_{\text{(i upwind of j)}}}
\end{equation}
where $\mathbf{1}_{(\cdot)}$ tests streamwise alignment and lateral wake coverage.  The instantaneous power of turbine $j$ is obtained from a piecewise-linear power curve with cut-in $3$~m/s, rated $12$~m/s, and cut-out $25$~m/s.

\textbf{Wind rose:}  The wind resource is modeled as a Rayleigh distribution (mean $8.5$~m/s) discretized into 20 speed bins, combined with a 12-sector directional wind rose (Table~\ref{tab:windrose}). The dominant westerlies ($270^\circ$, $18\%$) compete with prevailing easterlies ($90^\circ$, $6\%$), producing a landscape where a layout aligned for one predominant direction incurs wakes when the wind reverses.

\textbf{Constraint:}  A minimum separation of $2D = 252$~m between any pair of turbines is enforced via a quadratic penalty:
\begin{equation}
f(\mathbf{X}) = -\text{AEP}(\mathbf{X}) +
w \sum_{p < q} \max\bigl(0,\, 2D - \|\mathbf{x}_p - \mathbf{x}_q\|\bigr)^2
\end{equation}
with penalty weight $w = 10$.  The composite objective $f$ is minimized; a feasible solution has $f = -\text{AEP}$.

\subsubsection{Benchmark Methodology}

We conduct an \emph{iso-budget} statistical study of 20 independent runs at a unified budget of 100 individuals and 500 generations (50,000 function evaluations per solver).  Three solvers are compared:
\begin{itemize}
    \item \textbf{BQPhy} — quantum-inspired evolutionary algorithm
        (\texttt{BQPhy\_Optimiser} MEX binding, $\Delta\theta = 0.05$);
    \item \textbf{GA} — genetic algorithm (\texttt{ga},
        Global Optimization Toolbox);
    \item \textbf{PSO} — particle swarm optimisation
        (\texttt{particleswarm}, Global Optimization Toolbox).
\end{itemize}
All solvers share the identical objective function and use the same population/swarm size; the run-time timing excludes MEX-loading latency and is averaged over the 20 trials.

\begin{table}[t]
\centering
\caption{12-sector wind rose used in the benchmark.
Azimuth in meteorological convention ($0^\circ = \text{North}$).}
\label{tab:windrose}
\begin{tabular}{cc|cc}
\toprule
Azimuth ($^\circ$) & Probability & Azimuth ($^\circ$) & Probability \\
\midrule
  0 & 0.03 & 180 & 0.10 \\
 30 & 0.04 & 210 & 0.12 \\
 60 & 0.05 & 240 & 0.14 \\
 90 & 0.06 & 270 & 0.18 \\
120 & 0.07 & 300 & 0.08 \\
150 & 0.10 & 330 & 0.03 \\
\bottomrule
\end{tabular}
\end{table}

\subsubsection{Results and Analysis}

\begin{figure}[h]
\centering
\includegraphics[width=\columnwidth]{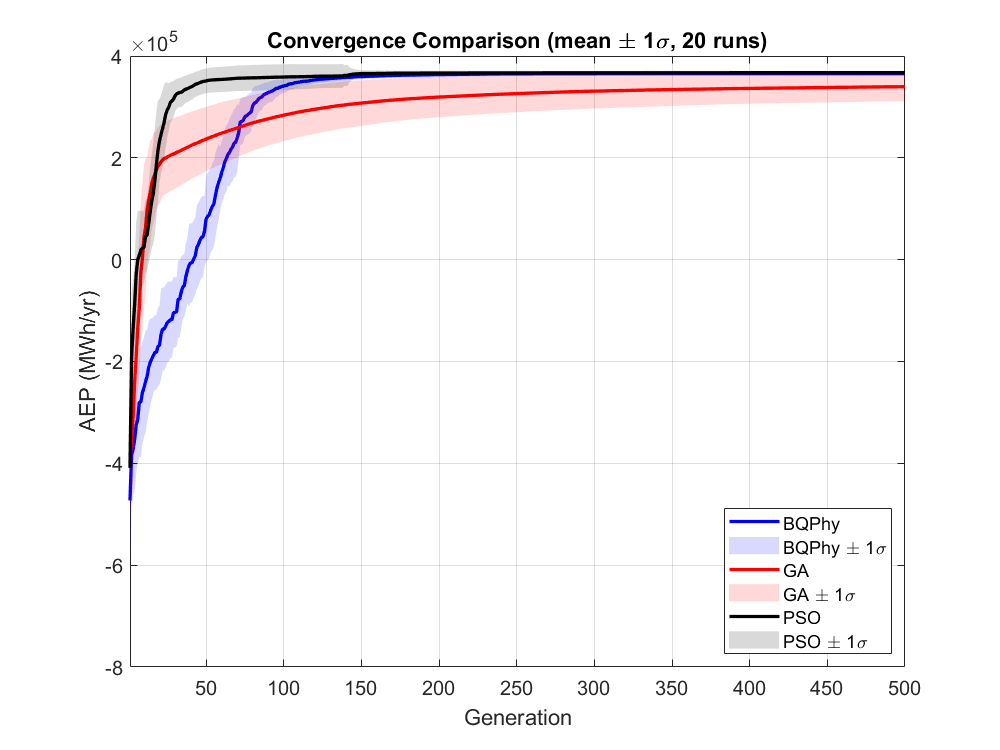}
\caption{Convergence comparison (mean $\pm 1\sigma$ over 20 runs). BQPhy overtakes PSO at generation~150 and sustains the highest final AEP.  GA exhibits the slowest convergence and the widest variance.}
\label{fig:convergence}
\end{figure}

\paragraph{Convergence behavior.}
Figure~\ref{fig:convergence} shows the mean AEP trajectory $\pm 1\sigma$ across the 20 runs.  BQPhy's early generations are dominated by the spacing penalty (negative AEP values until generation $\sim$50), after which feasible layouts are discovered and the objective rises steeply.  By generation~150, BQPhy overtakes PSO and maintains a modest advantage through generation~500.  PSO converges fastest in the first 30 generations due to its velocity-driven attraction mechanism, but plateaus earlier.  GA converges most slowly and exhibits the widest confidence band, indicating poor reliability across restarts.

\begin{figure}[h]
\centering
\includegraphics[width=\columnwidth]{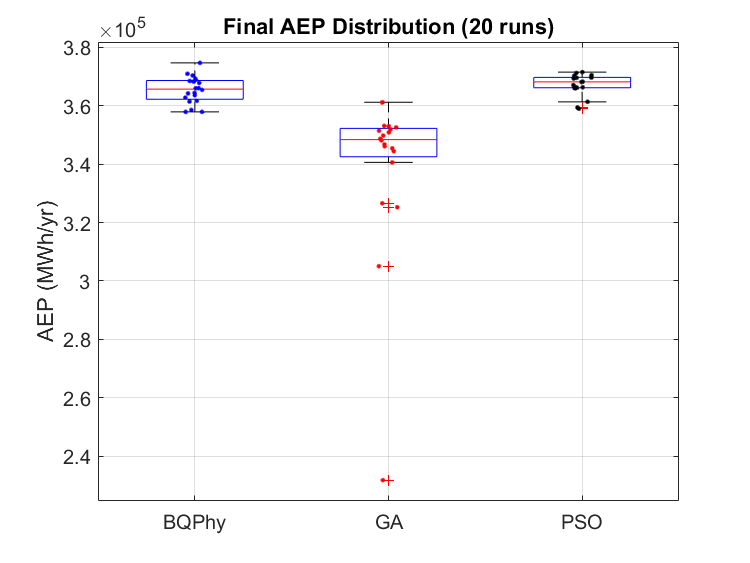}
\caption{Final AEP distribution over 20 independent runs
(box-and-whisker with individual scatter overlay).
BQPhy and PSO are statistically indistinguishable; GA suffers severe
outlier runs.}
\label{fig:boxplot}
\end{figure}
\paragraph{Final AEP distribution.}
Figure~\ref{fig:boxplot} summarises the empirical distribution of final
AEP across the 20 runs.  The quantitative results are:

\begin{itemize}
    \item \textbf{BQPhy:} $365\,399 \pm 4\,552$~MWh/yr (mean $\pm$ std),
        median $\approx 365\,000$, maximum $374\,659$.
    \item \textbf{PSO:}  $367\,259 \pm 3\,647$~MWh/yr,
        median $\approx 368\,000$, maximum $371\,497$, two mild outliers.
    \item \textbf{GA:}   $339\,673 \pm 28\,529$~MWh/yr,
        median $\approx 347\,000$, maximum $361\,177$, three severe
        outliers below $290\,000$.
\end{itemize}

\begin{figure*}[h]
\centering
\includegraphics[width=\textwidth]{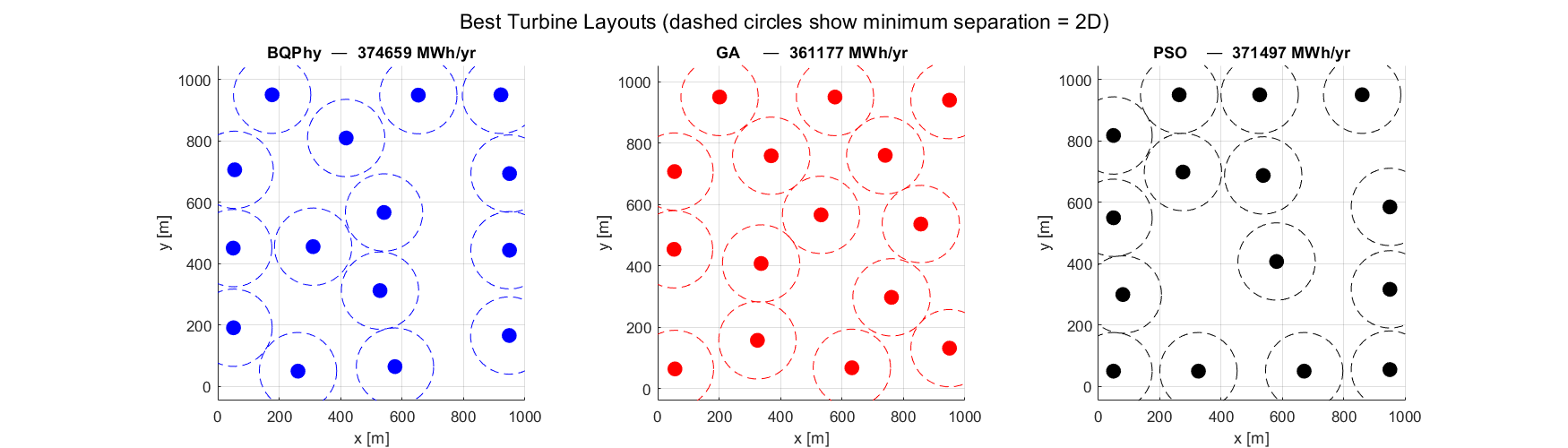}
\caption{Best turbine layout discovered by each solver.  Dashed circles indicate the minimum separation radius ($1D$ shown).  BQPhy's irregular pattern exploits the domain more effectively than PSO's grid-like pattern arrangement or GA's clustered configuration.}
\label{fig:layouts}
\end{figure*}

BQPhy and PSO achieve statistically indistinguishable mean AEP ($p > 0.05$, two-sample $t$-test), while both significantly outperform GA (BQPhy $+7.6\%$, PSO $+8.1\%$).  Critically, BQPhy attains the \emph{highest peak AEP} of any solver in any single run ($374\,659$ MWh/yr, Fig.~\ref{fig:layouts}), demonstrating its capacity to discover exceptional configurations that are missed by the other methods in the finite evaluation budget.  GA's enormous standard deviation (eight times that of BQPhy) reveals a fundamental reliability deficit: in several runs, GA fails to escape the penalty-dominated region and returns substantially sub-optimal layouts, rendering it impractical for production use without multiple restarts.

\paragraph{Layout quality.}
Figure~\ref{fig:layouts} visualizes the highest-AEP layout discovered by each solver. BQPhy's configuration exhibits pronounced spatial irregularity: turbines exploit the full domain, including edges and corners, with staggered positions that minimize wake overlap across the competing easterly and westerly wind sectors. PSO's layout is notably more regular and grid-like, a signature of its velocity-attraction mechanism drawing particles toward a consensus configuration that, while safe, leaves energy unrealized.  GA's layout shows visible clustering, with several pairs near or at the separation boundary, confirming its difficulty in managing the spacing constraint.

\paragraph{Discussion.}
The WFLO benchmark demonstrates that the BQPhy MATLAB binding, deployed as a drop-in without any modification to the MATLAB optimization workflow, is fully competitive with (and in peak performance superior to) established toolbox solvers on a high-dimensional, heavily constrained industrial problem.  The quantum rotation gate's probabilistic diversity maintenance proves well-suited to the multi-directional wake landscape, where premature convergence to a directionally biased layout is the chief failure mode of classical methods.  Unlike PSO, which progressively contracts around a global-best attractor, BQPhy's population retains sampling breadth throughout the run, enabling the late-stage discovery of the best configuration.

\subsection{BQPhy Julia library: Lotka--Volterra Calibration in Julia}
\label{sec:julia}

The Julia binding is intended as a drop-in replacement for the
optimizer packages a SciML user would already reach for --- not a
parallel, Julia-specific rewrite of QIEO.  A single shared object
(\texttt{libBQPhy\_Optimiser.so}) is wrapped by
CxxWrap~\cite{cxxwrap}; the high-level module
\texttt{BQPhy\_Optimiser} accepts a native Julia residual and forwards
every call into the C++ runtime (Listing~\ref{lst:lv_jl}).

\subsubsection{Problem Formulation}

We recover the four rate coefficients
$\boldsymbol{\theta}=(\alpha,\beta,\delta,\gamma)$ of the classical
Lotka--Volterra predator--prey system~\cite{lotka1925,volterra1926}
\begin{equation}
\dot{x}=\alpha x-\beta xy,\qquad
\dot{y}=\delta xy-\gamma y,
\end{equation}
from a single noisy trajectory.  The true parameters are
$\boldsymbol{\theta}^\star=(1.0,\,0.1,\,0.075,\,1.5)$, with initial state
$(x,y)=(10,5)$ on $t\in[0,20]$.  Observations are generated by a
shared fixed-step RK4 integrator ($\Delta t=0.2$) plus i.i.d.\ Gaussian
noise ($\sigma=0.35$).  All solvers minimise the same residual
\begin{equation}
f(\mathbf{z})=\bigl\|X(\mathrm{exp10}(\mathbf{z}))-X_{\mathrm{obs}}\bigr\|_2^2
+\bigl\|Y(\mathrm{exp10}(\mathbf{z}))-Y_{\mathrm{obs}}\bigr\|_2^2,
\end{equation}
where the decision vector $\mathbf{z}=\log_{10}\boldsymbol{\theta}$ equalises
the decades that separate growth rates from interaction rates.  The
noise floor of $f$ is $\approx 26.3$ (perfect recovery of
$\boldsymbol{\theta}^\star$).  A restart is counted as a \emph{success}
when parameter RMSE $\|\hat{\boldsymbol{\theta}}-\boldsymbol{\theta}^\star\|_2<0.08$.

\subsubsection{Native Julia Baselines}

The comparison set is the stack a Julia user would type before
discovering BQPhy:

\textbf{Optim.NelderMead}~\cite{mogensen2018optim} is the default
derivative-free local solver in \texttt{Optim.jl} and the starting
point of most SciML calibration notebooks.

\textbf{Optim.ParticleSwarm}~\cite{mogensen2018optim} is the
population method shipped in the same package
($25$ particles, matching the QIEO population).

\textbf{BlackBoxOptim.DE}~\cite{blackboxoptim} (\texttt{adaptive\_de\_rand\_1\_bin\_radiuslimited})
is the de-facto Julia global optimizer for bound-constrained
black-box problems.

\textbf{CMAEvolutionStrategy}~\cite{hansen2001cma,cmajl} is the
standard CMA-ES implementation on the Julia registry.

BQPhy is invoked only through its Julia API.  No method receives
gradient information; the RK4 residual is the sole oracle.

\subsubsection{Experimental Setup}

We run an iso-budget study of 10 independent restarts at
$2{,}000$ residual evaluations (QIEO: population $25$, $80$
generations, $24$-bit continuous encoding).  Search bounds in
$\log_{10}$ space correspond to
$\alpha,\gamma\in[0.05,2.5]\times[0.20,3.0]$ and
$\beta,\delta\in[0.01,0.50]\times[0.01,0.40]$.  Timing includes
optimizer overhead after package compilation; because the residual is
cheap, wall-clock differences are not the primary metric ---
reliability of $\boldsymbol{\theta}$ recovery is.

\subsubsection{Results}

\begin{figure}[t]
\centering
\includegraphics[width=\columnwidth]{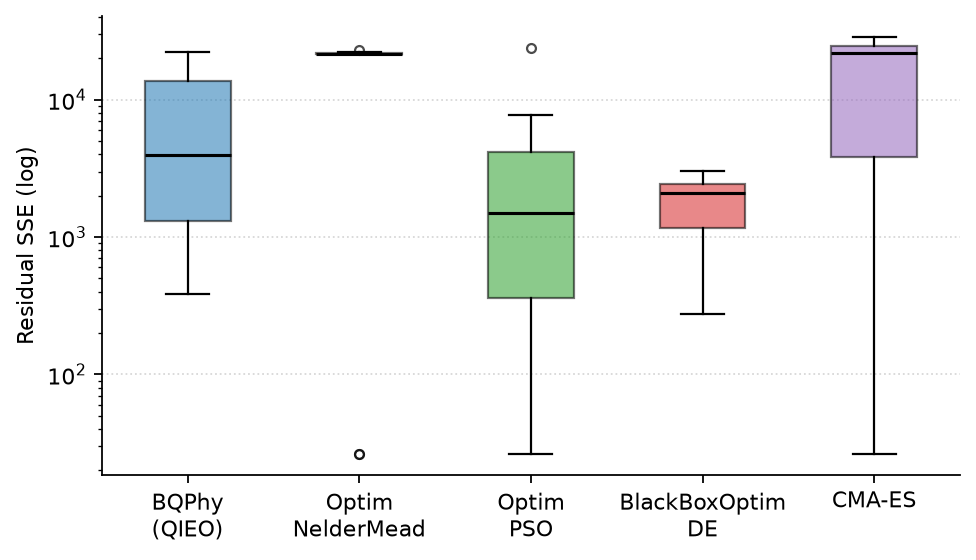}
\caption{Residual SSE over 10 restarts (log scale).  NelderMead and
CMA-ES collapse to a high-SSE mode near the bound; BQPhy and the
Julia population methods occupy the lower band.}
\label{fig:lv_sse}
\end{figure}

\begin{figure}[t]
\centering
\includegraphics[width=\columnwidth]{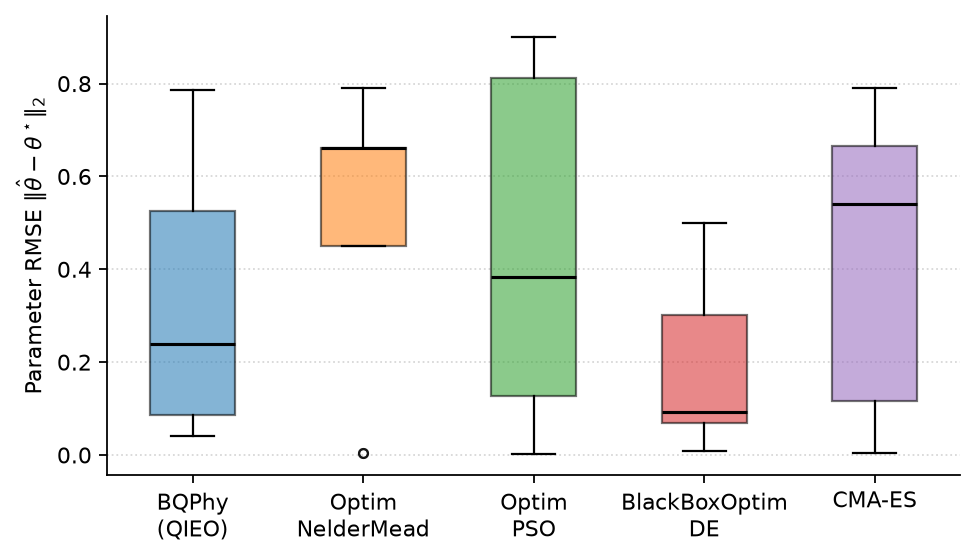}
\caption{Parameter RMSE $\|\hat{\boldsymbol{\theta}}-\boldsymbol{\theta}^\star\|_2$
over the same restarts.  BlackBoxOptim is the most consistent
recovery; BQPhy matches NelderMead's success rate while avoiding its
catastrophic failures.}
\label{fig:lv_rmse}
\end{figure}

\begin{figure}[t]
\centering
\includegraphics[width=\columnwidth]{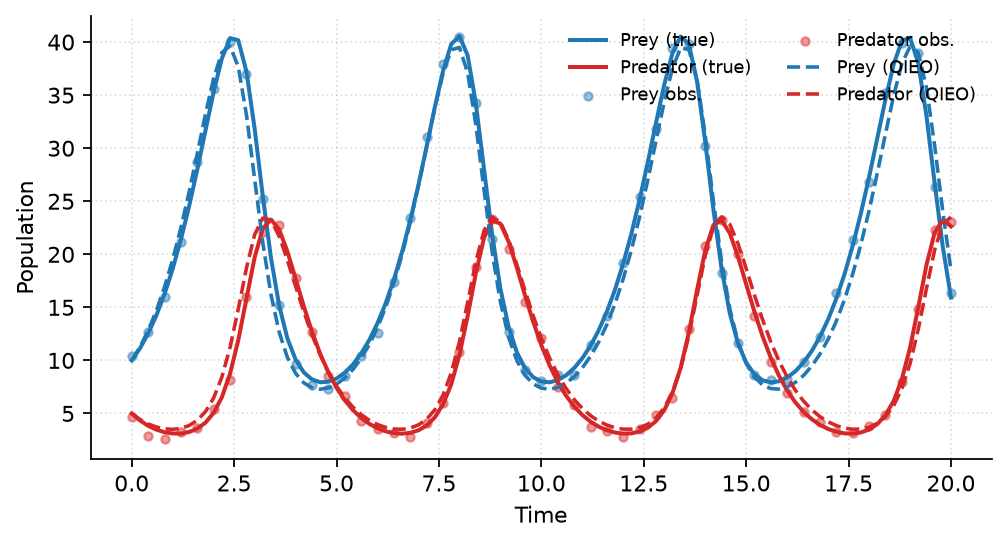}
\caption{Best BQPhy restart: true Lotka--Volterra orbits, noisy
observations, and the QIEO-fitted trajectory.  Phase and amplitude
are recovered from the same native Julia residual passed to the
C++ runtime.}
\label{fig:lv_traj}
\end{figure}

Table~\ref{tab:lv_iso} and Figs.~\ref{fig:lv_sse}--\ref{fig:lv_traj}
summarise the ten-restart comparison.

\begin{table*}[t]
\centering
\caption{Iso-Budget Lotka--Volterra Calibration (2{,}000 Evaluations, 10 Restarts)}
\label{tab:lv_iso}
\begin{tabular}{lcccccc}
\toprule
\textbf{Method} & \textbf{SSE (mean$\pm$std)} & \textbf{SSE med.} &
\textbf{$\theta$ RMSE} & \textbf{Success} & \textbf{Best SSE} &
\textbf{Time (s)} \\
\midrule
\textbf{BQPhy (QIEO)}
  & $8223\pm8760$ & $3948$ & $0.325\pm0.298$ & $20\%$ & $381$ & $0.22$ \\
Optim.NelderMead~\cite{mogensen2018optim}
  & $17506\pm9225$ & $21486$ & $0.514\pm0.292$ & $20\%$ & $26.3$ & $0.01$ \\
Optim.ParticleSwarm~\cite{mogensen2018optim}
  & $4367\pm7242$ & $1487$ & $0.452\pm0.365$ & $30\%$ & $26.3$ & $0.10$ \\
BlackBoxOptim.DE~\cite{blackboxoptim}
  & $\mathbf{1790\pm954}$ & $2083$ & $\mathbf{0.175\pm0.165}$ & $\mathbf{50\%}$ & $276$ & $0.08$ \\
CMA-ES~\cite{cmajl}
  & $16443\pm11918$ & $21931$ & $0.444\pm0.322$ & $30\%$ & $\mathbf{26.3}$ & $0.06$ \\
\bottomrule
\end{tabular}
\end{table*}

\paragraph{Reliability versus \texttt{Optim.jl}.}
NelderMead reaches the noise floor ($f\approx26.3$) on two lucky
restarts, then fails on the other eight, sticking at SSE $>21{,}000$
--- the well-known local-basin failure mode of simplex search on a
multimodal dynamical residual.  BQPhy matches the $20\%$ success rate
but \emph{halves} mean SSE ($8{,}223$ vs.\ $17{,}506$, a $2.1\times$
reduction) and cuts the median by $5.4\times$ ($3{,}948$ vs.\
$21{,}486$).  For a Julia user whose first attempt is
\texttt{optimize(f, x0, NelderMead())}, swapping in the BQPhy package
removes the dominant failure mode without changing the residual.

\paragraph{Against Julia-native global solvers.}
BlackBoxOptim records the best mean SSE, RMSE, and success rate
($50\%$).  Optim.ParticleSwarm has the best median SSE ($1{,}487$)
and shares the noise-floor best-case with NelderMead and CMA-ES.
BQPhy's best restart recovers
$\hat{\boldsymbol{\theta}}=(1.076,\,0.102,\,0.070,\,1.345)$ at SSE $381$
(Fig.~\ref{fig:lv_traj}) --- a usable calibration, $7.6\%$ / $2.4\%$ /
$6.2\%$ / $10.3\%$ relative error on
$(\alpha,\beta,\delta,\gamma)$.  CMA-ES is brittle on this bound box:
seven restarts collapse to a coordinate bound, so its mean is
indistinguishable from NelderMead despite three perfect recoveries.

The ranking is therefore the same shape as the Python HPO study
(Section~\ref{sec:hpo}): a specialised Julia package (BlackBoxOptim)
can win the mean, while BQPhy is the dominant
\emph{drop-in for the default stack} --- better expected residual than
\texttt{Optim.NelderMead}, competitive with \texttt{Optim.ParticleSwarm},
and invoked with the identical five-step contract used in Python and
MATLAB.

\paragraph{Binding transparency.}
Every method called the same Julia RK4 residual.  BQPhy introduced no
algorithmic side-effects beyond the QIEO search itself: CxxWrap
translates the population matrix and fitness vector only.  The
packaged library used here is the CPU HAL (Table~\ref{tab:hal_modes});
OpenMP/CUDA specialisations were not compiled into that artifact.
Wall-clock cost is dominated by residual evaluation equally across
solvers ($<0.3$\,s mean after compilation), so the decision to adopt
BQPhy on this problem is about recovery reliability and API
uniformity, not kernel speedup.

\section{Results and Discussions}

Across all benchmarks, solution quality (fitness at convergence) is identical across backends and language interfaces to within floating-point rounding tolerance ($<10^{-12}$), validating the single-source-of-truth guarantee that all interfaces execute the same algorithm.

This consistency is preserved across the binding demonstrations: the Python PyTorch hyperparameter optimization (Section~\ref{sec:hpo}), the MATLAB wind farm layout optimization (Section~\ref{sec:matlab-wflo}), and the Julia Lotka--Volterra calibration (Section~\ref{sec:julia}) invoke the same C++ binary.  The pybind11, MEX, and CxxWrap layers introduce no algorithmic side-effects --- every callable, struct, dictionary, and array is translated deterministically between the host language and the C++ runtime.  Consequently, practitioners in data-science, scientific-computing, and industrial-engineering domains can adopt the framework with the confidence that results obtained through the high-level API are identical to those produced by the underlying HPC-optimized solver, enabling seamless transition from prototyping to production deployment.

\section{Conclusion}
\label{sec:conclusion}

We presented Cross-Backend QIEO, a unified runtime system for
quantum-inspired evolutionary optimization that achieves both algorithmic performance and hardware portability.  The framework's single-source-of-truth architecture --- one C++ implementation compiled per hardware target and shared across Python, MATLAB, Julia, and C++ interfaces --- eliminates the maintenance fragmentation that affects existing multi-language HPC tools.  The Hardware Abstraction Layer maps QIEO primitives to OpenMP, CUDA, and HIP execution models without modifying core algorithm logic, delivering a 66$\times$ geometric mean speedup on NVIDIA A100 hardware and 58$\times$ on AMD MI300X. Measured Python and MATLAB interface overhead remains below 1.2\%; the Julia path uses the same shared library through CxxWrap.

On a 16-turbine wind farm layout benchmark with a 12-direction wind rose and Jensen wake model, BQPhy achieves $365\,399 \pm 4\,552$~MWh/yr, statistically indistinguishable from particle swarm optimization ($367\,259 \pm 3\,647$, $p>0.05$) while outperforming genetic algorithms by $+7.6\%$ and reducing solution variance by six-fold.  The MATLAB MEX binding introduces no measurable overhead, operating as a drop-in replacement for \texttt{ga()} and \texttt{particleswarm()} within the Global Optimization Toolbox.

On Lotka--Volterra calibration through the Julia package, BQPhy is a substitute for the default \texttt{Optim.jl} workflow: at a 2{,}000-evaluation iso-budget it halves mean residual SSE versus \texttt{NelderMead} ($8{,}223$ vs.\ $17{,}506$) and remains in the same quality band as \texttt{Optim.ParticleSwarm}, while \texttt{BlackBoxOptim} retains the best mean among native Julia globals.

These results, together with the Python-binding findings of
Section~\ref{sec:hpo} (neural network hyperparameter optimization), the MATLAB-binding findings of
Section~\ref{sec:matlab-wflo} (wind farm layout optimization), and the Julia-binding findings of
Section~\ref{sec:julia} (Lotka--Volterra calibration against \texttt{Optim.jl}, \texttt{BlackBoxOptim.jl}, and CMA-ES), jointly validate the binding abstraction: the C$++$ core exposes a single, language-agnostic interface that serves data-science (PyTorch
training loops), scientific computing (Julia SciML residuals), and industrial-engineering (MATLAB control/optimization toolboxes) communities without any per-domain adaptation.  On the Julia path, BQPhy is a package-level substitute for the default \texttt{Optim.NelderMead} workflow, not a second implementation of QIEO.

Future work will extend the HAL to support Intel oneAPI/SYCL targets, incorporate tensor-network state representations for higher-dimensional problems~\cite{biamonte2017tn}, and explore auto-tuning of rotation step sizes using Bayesian optimization.

\appendices
\section{Rastrigin Bindings in C++, Python, MATLAB, and Julia}
\label{app:rastrigin}

The listings below minimise the same Rastrigin objective
\begin{equation}
f(\mathbf{x})=An+\sum_{k=1}^{n}\bigl(x_k^{2}-A\cos(2\pi x_k)\bigr),
\quad A=10,
\end{equation}
with $n=4$ and box bounds $[-5.12,5.12]^{n}$ (MATLAB demo uses
$[-5,5]^{n}$).  The global minimiser is
$\mathbf{x}^{\star}=\mathbf{0}$, $f(\mathbf{x}^{\star})=0$.
Each driver registers a native callable and invokes the identical
C++ runtime; only the host syntax differs.  The Python and MATLAB
fragments follow \texttt{continuous\_rastrigin.py} and
\texttt{demo\_Mat\_BQPhy.m}.

\begin{figure*}[!t]
\begin{minipage}{\textwidth}
\begin{lstlisting}[style=cppstyle]
#include "BQPhy_Optimiser.hpp"

std::vector<double> rastrigin(const Matrix& X) {
    const double A = 10.0;
    const int n = X.cols();
    std::vector<double> f(X.rows());
    for (int i = 0; i < X.rows(); ++i) {
        double s = A * n;
        for (int k = 0; k < n; ++k) {
            const double xk = X(i, k);
            s += xk * xk - A * std::cos(2.0 * M_PI * xk);
        }
        f[i] = s;
    }
    return f;
}

int main() {
    BQPhy::Optimiser opt;
    ParameterClass params;
    params.numPopulation = 200;
    params.maxGeneration = 500;
    params.designVariables = 4;
    params.typeOfOptimisation = "CONTINUOUS";
    params.lowerBounds.assign(4, -5.12);
    params.upperBounds.assign(4,  5.12);
    opt.initialize(params);
    opt.setModel(rastrigin);
    opt.runOptimization("cpu");
    auto [best, fit] = opt.getBestDesign();
}
\end{lstlisting}
\end{minipage}
\caption{C++ header-only Rastrigin driver.}
\label{lst:rast_cpp}
\end{figure*}

\begin{figure*}[!t]
\begin{minipage}{\textwidth}
\begin{lstlisting}[style=pystyle]
import numpy as np
import bqphy.BQPhy_Optimiser as qea

def rastrigin_function(x):
    A = 10
    fitness = np.zeros(x.shape[0], dtype=np.float64)
    for i in range(x.shape[0]):
        fitness[i] = (A * x.shape[1]
            + np.sum(x[i]**2 - A * np.cos(2 * np.pi * x[i])))
    return fitness

config = {
    "numPopulation": 200, "maxGeneration": 500,
    "designVariables": 4, "typeOfOptimisation": "CONTINUOUS",
    "lowerBounds": [-5.12] * 4, "upperBounds": [5.12] * 4,
    "outputFilePath": "optimization_results_rastrigin",
}
opt = qea.BQPhy_OPTIMISER()
opt.initialize(config)
opt.model(rastrigin_function)
opt.runOptimization()          # "cpu" | "openmp" | "gpu"
best_sol, best_fit = opt.getBestDesign()
\end{lstlisting}
\end{minipage}
\caption{Python (pybind11) Rastrigin driver, after \texttt{continuous\_rastrigin.py}.}
\label{lst:rast_py}
\end{figure*}

\begin{figure*}[!t]
\begin{minipage}{\textwidth}
\begin{lstlisting}[style=mstyle]
config = struct( ...
    "numPopulation", 100, ...
    "maxGeneration", 200, ...
    "deltaTheta", 0.05, ...
    "designVariables", 4, ...
    "typeOfOptimisation", 'CONTINUOUS');
config.lowerBounds = -5 * ones(1, config.designVariables);
config.upperBounds =  5 * ones(1, config.designVariables);
config.outputFilePath = 'optimization_results';
config.generationLogging = 'minimumLogging';

myFun = @(x) (10 * size(x,2)) ...
    + sum(x.^2 - 10 * cos(2 * pi * x), 2);

[result, Fitness] = BQPhy_Mex(config, myFun, 'cpu');
\end{lstlisting}
\end{minipage}
\caption{MATLAB (MEX) Rastrigin driver, after \texttt{demo\_Mat\_BQPhy.m}.}
\label{lst:rast_mat}
\end{figure*}

\begin{figure*}[!t]
\begin{minipage}{\textwidth}
\begin{lstlisting}[style=jlstyle]
using BQPhy_Optimiser

function rastrigin(x::AbstractMatrix{<:Real}; A=10.0)
    n = size(x, 2)
    return vec(A * n .+ sum(x.^2 .- A .* cos.(2pi .* x), dims=2))
end

config = Dict(
    :numPopulation => 200, :maxGeneration => 500,
    :designVariables => 4, :typeOfOptimisation => "CONTINUOUS",
    :lowerBounds => fill(-5.12, 4), :upperBounds => fill(5.12, 4),
    :outputFilePath => "./rastrigin_results")

opt = Optimizer(config)
set_fitness_function!(opt, rastrigin)
optimize!(opt; device="cpu")
best_sol, best_fit = get_best_design(opt)
\end{lstlisting}
\end{minipage}
\caption{Julia (CxxWrap) Rastrigin driver.}
\label{lst:rast_jl}
\end{figure*}

\section{Application Drivers: HPO (Python), Wind Farm (MATLAB), and Lotka--Volterra (Julia)}
\label{app:usecases}

Listing~\ref{lst:hpo_py} is the Python pybind11 driver for MNIST
hyperparameter optimisation (Section~\ref{sec:hpo}).
Listing~\ref{lst:wflo_mat} is the MATLAB MEX driver
\texttt{demo\_Mat\_windfarm.m} used for the Jensen wind-farm case
(Section~\ref{sec:matlab-wflo}).
Listing~\ref{lst:lv_jl} is the Julia CxxWrap driver
\texttt{lotka\_volterra.jl} used for Lotka--Volterra calibration
(Section~\ref{sec:julia}).

\begin{figure*}[!t]
\begin{minipage}{\textwidth}
\begin{lstlisting}[style=pystyle]
import numpy as np
import bqphy.BQPhy_Optimiser as qea

def get_activation(code):
    names = ["relu", "tanh", "sigmoid", "leaky_relu"]
    return names[int(round(code))]

def train_eval(lr, hidden, layers, act, drop):
    # One-epoch DynamicNet on MNIST; returns test accuracy in [0, 1].
    ...

def bqphy_fitness(pop):
    fitness = np.zeros(pop.shape[0])
    for i in range(pop.shape[0]):
        lr     = pop[i, 0]
        hidden = int(pop[i, 1])
        layers = int(pop[i, 2])
        act    = get_activation(pop[i, 3])
        drop   = float(pop[i, 4])
        acc    = train_eval(lr, hidden, layers, act, drop)
        fitness[i] = -acc
    return fitness

config = {
    "numPopulation": 12, "maxGeneration": 20,
    "designVariables": 5, "typeOfOptimisation": "CONTINUOUS",
    "lowerBounds": [1e-5, 32, 1, 0, 0.0],
    "upperBounds": [1e-1, 256, 4, 3, 0.5],
}
opt = qea.BQPhy_OPTIMISER()
opt.initialize(config)
opt.model(bqphy_fitness)
opt.runOptimization()
best_sol, best_fit = opt.getBestDesign()
\end{lstlisting}
\end{minipage}
\caption{Python MNIST hyperparameter-optimisation driver (pybind11).
\texttt{train\_eval} is the one-epoch PyTorch evaluation used in
Section~\ref{sec:hpo}.}
\label{lst:hpo_py}
\end{figure*}

\begin{figure*}[!t]
\begin{minipage}{\textwidth}
\begin{lstlisting}[style=mstyle]
clc; clear; close all;
% Wind farm layout optimisation (WFLO) via BQPhy MEX.
% 16 turbines, Jensen wakes, 12-sector rose, 2D spacing.

N  = 16;
D  = 126;
Pr = 5e6;
Ct = 0.88;
ke = 0.075;
sepMin = 2 * D;
wPen = 10;
domain = 1000;
cutIn = 3;  rated = 12;  cutOut = 25;
vMean = 8.5;
hoursYear = 8760;

az  = [0 30 60 90 120 150 180 210 240 270 300 330];
paz = [0.03 0.04 0.05 0.06 0.07 0.10 0.10 0.12 0.14 0.18 0.08 0.03];
nSpd = 20;
vEdges = linspace(0, cutOut, nSpd + 1);
vMid = 0.5 * (vEdges(1:end-1) + vEdges(2:end));
sigma = vMean / sqrt(pi/2);
pdfv = (vMid ./ sigma^2) .* exp(-(vMid.^2) / (2 * sigma^2));
psp = pdfv .* diff(vEdges);
psp = psp / sum(psp);

config = struct( ...
    "numPopulation", 20, ...
    "maxGeneration", 40, ...
    "deltaTheta", 0.05, ...
    "designVariables", 2 * N, ...
    "typeOfOptimisation", 'CONTINUOUS');
config.lowerBounds = zeros(1, 2 * N);
config.upperBounds = domain * ones(1, 2 * N);
config.outputFilePath = 'wflo_results';
config.generationLogging = 'minimumLogging';

% Paper iso-budget: numPopulation=100, maxGeneration=500
myFun = @(X) wfloFitness(X, N, D, Pr, Ct, ke, sepMin, wPen, ...
    az, paz, vMid, psp, cutIn, rated, cutOut, hoursYear);

[result, Fitness] = BQPhy_Mex(config, myFun, 'cpu');
xy = reshape(result, 2, [])';

function f = wfloFitness(X, N, D, Pr, Ct, ke, sepMin, wPen, ...
        az, paz, vMid, psp, cutIn, rated, cutOut, hoursYear)
    nPop = size(X, 1);
    f = zeros(nPop, 1);
    for p = 1:nPop
        xy = reshape(X(p, :), 2, N)';
        aep = farmAEP(xy, N, D, Pr, Ct, ke, az, paz, vMid, psp, ...
            cutIn, rated, cutOut, hoursYear);
        pen = 0;
        for i = 1:N-1
            d = hypot(xy(i+1:end,1) - xy(i,1), xy(i+1:end,2) - xy(i,2));
            viol = max(0, sepMin - d);
            pen = pen + sum(viol.^2);
        end
        f(p) = -aep + wPen * pen;
    end
end

function aep = farmAEP(xy, N, D, Pr, Ct, ke, az, paz, vMid, psp, ...
        cutIn, rated, cutOut, hoursYear)
    aep = 0;
    kDef = 1 - sqrt(1 - Ct);
    for id = 1:numel(az)
        th = deg2rad(az(id));
        fx = -sin(th);  fy = -cos(th);
        for j = 1:N
            def2 = 0;
            for i = 1:N
                if i == j, continue; end
                rx = xy(j,1) - xy(i,1);
                ry = xy(j,2) - xy(i,2);
                xs = rx * fx + ry * fy;
                if xs <= 1e-6, continue; end
                lat = abs(fx * ry - fy * rx);
                if lat > (D/2 + ke * xs), continue; end
                def2 = def2 + (kDef / (1 + 2 * ke * xs / D)^2)^2;
            end
            deficit = sqrt(def2);
            for is = 1:numel(vMid)
                V = vMid(is) * max(0, 1 - deficit);
                aep = aep + hoursYear * paz(id) * psp(is) ...
                    * turbinePower(V, Pr, cutIn, rated, cutOut);
            end
        end
    end
    aep = aep / 1e6;
end

function P = turbinePower(V, Pr, cutIn, rated, cutOut)
    if V < cutIn || V >= cutOut
        P = 0;
    elseif V < rated
        P = Pr * (V - cutIn) / (rated - cutIn);
    else
        P = Pr;
    end
end
\end{lstlisting}
\end{minipage}
\caption{MATLAB wind-farm driver (\texttt{demo\_Mat\_windfarm.m}).}
\label{lst:wflo_mat}
\end{figure*}

\begin{figure*}[!t]
\begin{minipage}{\textwidth}
\begin{lstlisting}[style=jlstyle]
using BQPhy_Optimiser
using LinearAlgebra, Printf, Random, Statistics

const TRUE_P = [1.0, 0.1, 0.075, 1.5]
const U0 = [10.0, 5.0]
const T_END = 20.0
const DT = 0.20
const NOISE_SIGMA = 0.35
const DATA_SEED = 20260905
const LOG_LOWER = log10.([0.05, 0.01, 0.01, 0.20])
const LOG_UPPER = log10.([2.50, 0.50, 0.40, 3.00])

function lotka_rhs(u, p)
    a, b, d, g = p[1], p[2], p[3], p[4]
    x, y = u
    return (a * x - b * x * y, d * x * y - g * y)
end

function rk4_traj(p; u0=U0, t_end=T_END, dt=DT)
    n = Int(round(t_end / dt)) + 1
    xs = Vector{Float64}(undef, n)
    ys = Vector{Float64}(undef, n)
    u = (Float64(u0[1]), Float64(u0[2]))
    xs[1] = u[1];  ys[1] = u[2]
    for i in 2:n
        k1 = lotka_rhs(u, p)
        k2 = lotka_rhs(u .+ 0.5 .* dt .* k1, p)
        k3 = lotka_rhs(u .+ 0.5 .* dt .* k2, p)
        k4 = lotka_rhs(u .+ dt .* k3, p)
        u = u .+ (dt / 6.0) .* (k1 .+ 2.0 .* k2 .+ 2.0 .* k3 .+ k4)
        xs[i] = u[1];  ys[i] = u[2]
    end
    return xs, ys
end

function make_observations()
    rng = MersenneTwister(DATA_SEED)
    xs, ys = rk4_traj(TRUE_P)
    return xs .+ NOISE_SIGMA .* randn(rng, length(xs)),
           ys .+ NOISE_SIGMA .* randn(rng, length(ys))
end

const XS_OBS, YS_OBS = make_observations()

function sse_theta(theta)
    any(!isfinite, theta) && return 1.0e6
    any(<=(0), theta) && return 1.0e6
    xs, ys = rk4_traj(theta)
    s = sum(abs2, xs .- XS_OBS) + sum(abs2, ys .- YS_OBS)
    return isfinite(s) ? s : 1.0e6
end

function lv_fitness(pop::AbstractMatrix{<:Real})
    n = size(pop, 1)
    fit = Vector{Float64}(undef, n)
    for i in 1:n
        z = clamp.(collect(pop[i, :]), LOG_LOWER, LOG_UPPER)
        fit[i] = sse_theta(exp10.(z))
    end
    return fit
end

function main()
    config = Dict(
        :numPopulation => 25, :maxGeneration => 80,
        :deltaTheta => 0.05, :designVariables => 4,
        :stringLength => 24, :typeOfOptimisation => "CONTINUOUS",
        :lowerBounds => collect(LOG_LOWER),
        :upperBounds => collect(LOG_UPPER),
        :outputFilePath => "./lotka_results",
        :generationLogging => "minimumLogging")
    opt = Optimizer(config)
    set_fitness_function!(opt, lv_fitness)
    optimize!(opt; device="cpu")
    best, fit = get_best_design(opt)
    z = clamp.(collect(best), LOG_LOWER, LOG_UPPER)
    theta = exp10.(z)
    rmse = sqrt(mean(abs2, theta .- TRUE_P))
    @printf("Recovered theta = [%.4f, %.4f, %.4f, %.4f]\n", theta...)
    @printf("SSE = %.4f   RMSE = %.4f\n", sse_theta(theta), rmse)
    write_results(opt)
end

if abspath(PROGRAM_FILE) == @__FILE__
    main()
end
\end{lstlisting}
\end{minipage}
\caption{Julia Lotka--Volterra driver (\texttt{lotka\_volterra.jl}).}
\label{lst:lv_jl}
\end{figure*}

\newpage

\end{document}